\documentclass[preprint,tightenlines,prd,showpacs]{revtex4}
\def\beq{\begin{equation}} \def\eeq{\end{equation}}
\def\bea{\begin{eqnarray}} \def\eea{\end{eqnarray}}

 \def\w{\Omega}
        \def\G{\Gamma}

        \def\H{{\cal H}}
        
        \def\S{{\cal S}}
        
        \def\c{{\cal C}}
        \def\d{{\rm d}}

\begin{document}

\title{Note on Canonical Quantization and \\
Unitary Equivalence in Field Theory}
\author{Alejandro Corichi}\email{corichi@nuclecu.unam.mx}
\affiliation{Instituto de Ciencias Nucleares\\
Universidad Nacional Aut\'onoma de M\'exico\\
A. Postal 70-543, M\'exico D.F. 04510, MEXICO}
\author{Jer\'onimo Cortez}\email{cortez@nuclecu.unam.mx}
\affiliation{Instituto de Ciencias Nucleares\\
Universidad Nacional Aut\'onoma de M\'exico\\
A. Postal 70-543, M\'exico D.F. 04510, MEXICO}
\author{Hernando Quevedo}\email{quevedo@nuclecu.unam.mx}
\affiliation{Instituto de Ciencias Nucleares\\
Universidad Nacional Aut\'onoma de M\'exico\\
A. Postal 70-543, M\'exico D.F. 04510, MEXICO}

\date{Feb 28th}

\begin{abstract}
The problem of defining and constructing representations of the
Canonical Commutation Relations can be systematically approached
via the technique of {\it algebraic quantization}. In particular,
when the phase space of the system is linear and finite
dimensional,  the `vertical polarization'  provides an unambiguous
quantization. For infinite dimensional field theory systems, where
the Stone-von Neumann theorem fails to be valid, even the simplest
representation, the Schr\"odinger functional picture has some
non-trivial subtleties. In this letter we consider the
quantization of a real free scalar field --where the Fock
quantization is well understood-- on an arbitrary background and
show that the representation coming from the most natural
application of the algebraic quantization approach is not, in
general, unitary equivalent to the corresponding
Schr\"odinger-Fock quantization. We comment on the possible
implications of this result for field quantization.
\end{abstract}

\maketitle

\section{Introduction}
\label{sec:1}

The process of constructing a quantum theory starting from a given
classical system, is not by any means a simple {\em recipe
procedure}. Indeed, the quantization process is full of choices
that might lead to inequivalent quantum theories. This ambiguity
in the quantization process is a well known fact, for instance,
when trying to define operators for quadratic momentum observables
due to factor ordering issues. In this letter we shall focus our
attention on canonical quantization methods; that is, we will
start by considering a phase space $\Gamma$ which, for simplicity,
can be linear with coordinates $(q^i,p_j)$. The final goal is to
have a quantum system described by a Hilbert space ${\cal H}$ and
a set of (Hermitian) operators $\hat{\cal O}_j$ associated with
real observables. In the past few decades, a manifold of methods
to deal with these problems have appeared with varying success. In
very broad terms, they differ on the basic objects/properties of
the classical system that is used to perform the quantization
process. For instance, the {\em group theoretical} quantization
places special attention on the symmetry group of the classical
phase space \cite{group}; the geometric quantization programme
tries to describe the process in terms of sections of bundles on
phase space \cite{wood}, etc. Of particular relevance is the
programme known as {\em algebraic quantization method}, pioneered
by Ashtekar \cite{AT} and its refined version (all tailored to
deal with constrained systems), known as refined algebraic
quantization \cite{RAQ}. The basic idea of this programme is to
identify a suitable algebra of observables on phase space for
which there will be unambiguous operators satisfying the canonical
commutation relations (CCR). The programme also involves the
choice of a vector space $V$ where the quantum operators
${\hat{\cal O}}_i$ are defined and an inner product on $V$ that
makes the operators have the desired {\em reality conditions}.
This constitutes the `kinematical' part of the quantization.
Dynamical issues, such as implementation of the Hamiltonian or
quantum constraints is generally regarded as a second step in the
process.


When the system is finite dimensional the quantization process has
a preferred endpoint in the well known Schr\"odinger
representation, which in the geometric quantization language,
represents the {\em vertical polarization} (wave functions depend
only on the configuration space variables $q^i$). For alternative
`polymer' representations where the configuration observable is
not well defined see \cite{polymer}. The Stone-von Neumann theorem
assures us that any `decent' representation will be unitarily
equivalent to the usual Schr\"odinger one.

In the case of a field theory system with an infinite number of
degrees of freedom, it is known that the CCR accept infinite
non-equivalent representations (for a very readable discussion on
this issue see \cite{wald}). This feature of field theory has an
important consequence in the definition of the QFT on curved
space-times where different observables might disagree on the
observable predictions of the theory. In the case of Fock
representations, the relevant construct in the quantization
process responsible for this ambiguity is well understood, in
terms of Bogoliubov transformations. Although several attempts
have been made to understand the subtleties of Schr\"odinger
representations \cite{jackiw,jemal}, a systematic construction for
general curved space-times has only been considered recently
\cite{ccq}. It is important then to explore and understand the
various objects involved in the quantization process and its
relation to the existence of (in-)equivalent quantum
representations. The purpose of this letter is to explore these
issues and analyze the algebraic quantization program in view of
the existing canonical quantization of the scalar field as done in
\cite{ccq}. That is, we shall follow the algebraic construction,
using the most natural choices along the way, and compare the
resulting quantization with \cite{ccq}.

The structure of the paper is as follows. In Sec.~\ref{sec:2} we
recall the algebraic quantization program in general and then
consider the case of the free real scalar field using this
prescription. In Sec.~\ref{sec:3}, we consider the quantization of
the scalar field, following the algebraic approach and then
compare this picture with the representation coming from a
different viewpoint \cite{ccq}. In Sec.~\ref{sec:4} we study the
origin of this discrepancy and recall the unitary implementation
of linear canonical transformation in quantum mechanics. We find
that even when naively, the two representations are unitarily
equivalent (as is the case in finite dimensional systems), the
required transformation does not exist; the two representation
are, in general, non-unitary. We conclude in Sec.~\ref{sec:5} with
a discussion of the results.

\section{Preliminaries}
\label{sec:2}

This section has two parts. In the first one, we recall basic features
about algebraic quantization and in the second part we review the
case of a real scalar field.

\subsection{Algebraic Quantization: General Formalism}

A physical system in the canonical perspective is normally
represented, at the classical level, by a {\it phase space}
$\Gamma$, endowed with a symplectic structure $\Omega$. The phase
space is denoted by $(\G,\Omega)$. The Lie algebra of vector
fields on $\G$ induces a Lie algebra structure on the space of
functions, given by $\{ f,g\} :=-\Omega_{ab} X^a_f X^b_g =
\Omega^{ab}\nabla_af\nabla_bg$ such that $X^a_{\{ f,g\}
}=-[X_f,X_g]^a$. The `product' $\{\cdot ,\cdot\}$ is called {\it
Poisson Bracket} (PB).

In very broad terms, by quantization one means the passage from a
classical system, as described in the last part, to a quantum
system. Observables on $\Gamma$ are to be promoted to self-adjoint
operators on a Hilbert space. However, we know that not all
observables can be promoted unambiguously to quantum operators
satisfying the CCR. A well known example of such problem is factor
ordering. What we {\it can} do is to construct a subset ${\cal S}$
of {\it elementary classical variables} for which the quantization
process has no ambiguity. This set ${\cal S}$ should satisfy two
properties:

{\it 1.} The set  ${\cal S}$ should be a vector space
large enough so that every
(regular) function on $\Gamma$ can be obtained by (possibly a limit of) sums
of products of elements in
${\cal S}$. The purpose of this condition is that we want that enough
observables are to be unambiguously quantized.

{\it 2.} The set ${\cal S}$ should be small enough such that it is closed
under Poisson brackets.

The next step is to construct an (abstract) quantum algebra ${\cal A}$ of
observables from the vector space ${\cal S}$ as the free associative
algebra generated by ${\cal S}$ (for a definition and discussion of
free associative algebras see \cite{geroch}). It is in this quantum
algebra ${\cal A}$ that we impose the Dirac quantization condition:
Given $A,B$ and $\{A,B\}$ in ${\cal S}$ we impose,
\beq
[\hat{A},\hat{B}]=i\hbar\widehat{\{A,B\}}\label{diracc}
\eeq
It is important to note that there is no factor order ambiguity in
the Dirac condition since $A,B$ and $\{A,B\}$ are contained in ${\cal S}$
and they have associated a unique element of ${\cal A}$.

The next step is to find a vector space $V$ and a representation
of the elements of ${\cal A}$ as operators on $V$. The reality
conditions (encoded as $*$-relations on ${\cal A}$) are used as a
criteria for finding the inner-product $\langle ,\rangle$ on $V$
that transforms these relations into Hermiticity conditions. That
is, real observables on ${\cal S}$ should become Hermitian
operators on $V$. Finally, one completes $V$ to get the Hilbert
space $\H$ of the theory. For details of this approach to
quantization see \cite{AT}.

In the case that the phase space $\Gamma$ is a linear space, there
is a particular simple choice for the set $\S$. We can take a
global chart on $\Gamma$ and we can choose $\S$ to be the vector
space generated by {\it linear} functions on $\Gamma$. In some
sense this is the smallest choice of $\S$ one can take. We can now
look at these linear functions on $\Gamma$. Denote by $Y^a$ an
element of $\Gamma$, and using the fact that it is linear space,
$Y^a$ also represents a vector in the tangent space $T\Gamma$.
Given a one form $\lambda_a$, we can define a linear function of
$\Gamma$ as follows: $F_\lambda(Y):=-\lambda_aY^a$. Note that
$\lambda$ is a label of the function with $Y^a$ as its argument.
First, note that there is a vector associated to $\lambda_a$:
\[
\lambda^a:=\Omega^{ab}\lambda_b
\]
so we can write
\beq
F_\lambda(Y)=\Omega_{ab}\lambda^aY^b=\Omega(\lambda,Y)
\eeq
If we are now given another label $\nu$, such that $G_\nu(Y)=\nu_aY^a$,
we can compute the Poisson Bracket
\beq
\{F_\lambda,G_\nu\}=\Omega^{ab}\nabla_aF_\lambda(Y)\nabla_bG_\nu(Y)=
\Omega^{ab}\lambda_a\nu_b
\eeq
Since the two-form is non-degenerate we can re-write it as
$\{F_\lambda,G_\nu\}=-\Omega_{ab}\lambda^a\nu^b$. Thus,
\beq
\{\Omega(\lambda,Y),\Omega(\nu,Y)\}=-\Omega(\lambda,\nu)
\eeq

 As a
concrete case, let us look at the example of a mechanical system
whose configuration space is $\c={\bf R}^3$. We can take a global
chart on $\Gamma$ given by $(q^i,p_i)$ and consider
$\S=\mbox{Span}\{1,q^1,q^2,q^3,p_1,p_2,p_3\}$. It is a seven
dimensional vector space. Notice that we have included the
constant functions on $\Gamma$, generated by the unit function
since we know that $\{q^i,p_j\}=\delta^i_j$, and we want $\S$ to
be closed under PB.

The quantum representation is the ordinary Schr\"odinger picture
where the Hilbert space is $\H=L^2({\bf R}^3,\d^3 x)$ and the
operators are represented by:
\begin{equation}
(\hat{1}\cdot\Psi)(q)=\Psi(q)\qquad
(\hat{q^i}\cdot\Psi)(q)=q^i\Psi(q)\qquad (\hat{p}_i\cdot\Psi)(q)=
-i\hbar\frac{\partial}{\partial q^i}\Psi(q)
\end{equation}
Thus, we recover the conventional Schr\"odinger
representation of the CCR.


\subsection{Real Scalar Field}

In this part we recall the classical theory of a real, linear
Klein-Gordon field $\phi$ with mass $m$ propagating on a
4-dimensional, globally hyperbolic spacetime
$(^{\mbox{\tiny{4}}\!}M,g_{ab})$. Global
hyperbolicity implies that $^{\mbox{\tiny{4}}\!}M$ has topology
${\bf{R}}\times \Sigma$,  and can be foliated by a one-parameter
family of smooth Cauchy surfaces diffeomorphic to $\Sigma$
with arbitrary embeddings of the
surface $\Sigma$ into $^{\mbox{\tiny{4}}\!}M$.

The phase space of the theory can be alternatively described by
the space $\G$ of Cauchy data (in the canonical approach), that
is, $\{(\varphi , \pi)\vert \, \varphi : \Sigma \to {\bf{R}},\,
\pi: \Sigma \to {\bf{R}}; \, \varphi ,
 \pi \in C^{\infty}_{0}(\Sigma)\}$ {\footnote{The class of functions
 comprised by Schwartz space is most commonly chosen for quantum
 field theory in Minkowski spacetime. However, the notion of Schwartz
  space is not extendible in any obvious way to more general manifolds
  \cite{wald,draft}. Hence, we shall define $\G$ to consist of initial
  data which are smooth and of compact support on $\Sigma$.}}, or by
  the space $S$ of smooth solutions to the Klein-Gordon equation which
  arises from initial data on $\G$ (in the covariant formalism) \cite{wald}.
  Note that, for each embedding $T_{t}:\Sigma \to {^{\mbox{\tiny{4}}\!}M}$,
  there exists an isomorphism ${\cal{I}}_{t}$ between $\G$ and $S$. The key
  observation is that there is a one to one correspondence between a pair
  of initial data of compact support on $\Sigma$, and solutions to the
  Klein-Gordon equation on $^{\mbox{\tiny{4}}\!}M$. That is to say:

Given an embedding $T_{t_{0}}$ of $\Sigma$ as a Cauchy surface
$T_{t_{0}}(\Sigma)$ in $^{\mbox{\tiny{4}}\!}M$, the (natural)
isomorphism ${\cal{I}}_{t_{0}}:\G\to S$ is obtained by taking a
point in $\G$ and evolving from the Cauchy surface $T_{t_{0}}(\Sigma)$
 to get a solution of $(g^{ab}\nabla_{a}\nabla_{b}-m^{2})\phi =0$.
 That is, the specification of a point in $\G$ is the appropriate initial
 data for determining a solution to the equation of motion. The inverse
map, ${\cal{I}}_{t_{0}}^{-1}:S \to \G$, takes a point $\phi \in S$
and finds the Cauchy data induced on $\Sigma$ by virtue of the
embedding $T_{t_{0}}$: $\varphi=T_{t_{0}}^{*}\phi$ and
$\pi=T_{t_{0}}^{*}(\sqrt{h}\pounds_{n}\phi)$, where $\pounds_{n}$
is the Lie derivative along the normal to the Cauchy surface
$T_{t_{0}}(\Sigma)$ and $h$ is the determinant of the induced
metric on such a surface. Note that the phase space $\G$ is of the
form $T^{*}{\c}$, where the classical configuration space ${\c}$
is comprised by the set of smooth real functions of compact
support on $\Sigma$.

Since the phase space $\G$ is a linear space, there is a
particular simple choice for the set of fundamental observables,
namely the vector space generated by {\it linear} functions on
$\Gamma$. More precisely, classical observables for the space $\G$
can be constructed directly by giving smearing functions on
$\Sigma$. We can define linear functions on $\G$ as follows: given
a vector $Y^a$ in $\G$ of the form $Y^a=(\varphi,\pi)^a$ (note
that due to the linear nature of $\G$, $Y^{a}$ also represents a
vector in $T \G$) and a pair $\lambda_a=(-f,-g)_a$, where $f$ is a
scalar density and $g$ a scalar, we define the action of $\lambda$
on $Y$ as, \beq \label{observ} F_\lambda(Y)=- \lambda_a
Y^a:=\int_\Sigma(f\varphi+g\pi)\,\d^3x \, . \eeq
 Now, since in the
phase space $\G$ the symplectic structure $\Omega$ takes the
following form, when acting on vectors $(\varphi_1,\pi_1)$
and$(\varphi_2,\pi_2)$, \beq
\Omega([\varphi_1,\pi_1],[\varphi_2,\pi_2])=\int_\Sigma(\pi_1\varphi_2-
\pi_2\varphi_1)\,\d^3x \, ,\label{symp2} \eeq then we can write
the linear function (\ref{observ}) in the form
$F_\lambda(Y)=\Omega_{ab}\lambda^a Y^b=\Omega(\lambda,Y)$, if we
identify $\lambda^b=\Omega^{ba}\lambda_a=(-g,f)^b$. That is, the
smearing functions $f$ and $g$ that appear in the definition of
the observables $F$ and are therefore naturally viewed as a 1-form
on phase space, can also be seen as the vector $(-g,f)^b$. Note
that the role of the smearing functions is interchanged in the
passing from a 1-form to a vector. Of particular importance for
what follows is to consider {\it configuration} and {\it momentum}
observables. They are particular cases of the observables $F$
depending of specific choices for the label $\lambda$. Let us
consider the ``label vector''
 $\lambda^a=(0,f)^a$, which would be normally regarded
as a vector in the ``momentum'' direction. However, when we
consider the linear observable that this vector generates, we get,
\begin{equation}
\varphi[f]:=\int_\Sigma \d^3\! x\,f\,\varphi\, .\label{observ2}
\end{equation}
Similarly, given the vector  $(-g,0)^a$ we can construct,
\begin{equation}
\pi[g]:=\int_\Sigma\d^3\! x\,g\,\pi .\label{observ3}
\end{equation}

Note that any pair of test fields $(-g,f)^a\in \G$ defines a
linear observable, but they are `mixed'. More precisely, a scalar
$g$ in $\Sigma$, that is, a pair $(-g,0)\in \G$ gives rise to a
{\it momentum} observable $\pi[g]$ and, conversely, a pair
$(0,f)^a$ yields a configuration observable.

\section{Quantum Scalar Field}
\label {sec:3}

This section has two parts. In the first one, we follow the
algebraic quantization procedure described in the last section,
for the case of a Gaussian (Fock) measure. In the second part we
recall the approach of \cite{ccq}, in which a different
representation of the CCR is obtained.

{\it Algebraic Quantization.} The algebraic quantization procedure
for the scalar field, in analogy with the finite dimensional
system, starts by considering the set ${\cal S}$ to be the span of
the identity and the functions $F_\lambda(Y)$. The vector space
$V$ where the abstract operators are to be represented will be
taken to be functionals of the configuration variable $\varphi$,
that is, elements of the form $\Psi[\varphi]$. In analogy with
quantum mechanics, the most natural representation of the basic
operators, when acting on functionals $\Psi[\varphi]$, is as
follows \beq
(\hat{\varphi}[f]\cdot\Psi)[\varphi]:=\varphi[f]\,\Psi[\varphi]\,
, \label{comfop} \eeq and \beq
(\hat{\pi}[g]\cdot\Psi)[\varphi]:=-i\hbar\int_{\Sigma}
\d^3\!x\;g(x)
 {{\delta \Psi}\over{\delta \varphi(x)}} + {\rm
multiplicative\; term}\, .\label{momop}
\eeq

The second term in (\ref{momop}), depending only on configuration
variable, is precisely there to render the operator self-adjoint
when the measure is different from the `homogeneous´ measure, and
it usually depends on the details of the measure. That is, given
the quantum measure, that for free field theories is known to be
``Gaussian'', one should adjust the action of the momentum
operator in order to satisfy the reality conditions. In what
follows we shall proceed with the choices that are motivated by
geometrical considerations. We know that if the momentum
observable can be associated to a vector field $v^a$ on
configuration space, then the general form of the momentum
operator is given by $\hat{P}(v)=-i\hbar(\pounds_v+\frac{1}{2}
{\rm Div}_\mu v)$, where ${\rm Div}_\mu v$ is the divergence of
the vector field $v$ with respect to the (quantum) measure $\mu$
(recall that a volume element is sufficient to define the
divergence of a vector field). Therefore, given the quantum
measure in the field theory case, one can in principle determine
the multiplicative term in the representation of the quantum
momentum operator.

Let us now recall the general form of the quantum measure in the
Schr\"odinger representation \cite{ccq}. Let $G$ by a positive,
Hermitian operator in the $L^2$ Hilbert space of scalar functions
$\varphi$. Then the general Gaussian measure is heuristically of
the form,
 \beq
 \d\mu_{\rm G}=\exp\left[ -\int_\Sigma
\d^3\!x\;\varphi\,G\,\varphi \right] {\cal D}\varphi \, , \eeq
where  ${\cal D}\varphi$ is the fictitious homogeneous functional
measure. Now, the vector fields on configuration space ${\cal C}$,
associated to the momentum variables $\pi[g]$ are constant vector
fields (in the chart defined by $(\varphi,\pi)$). There is a
general formula for the divergence of a `constant' vector field
$v^a$ given by ${\rm Div}_\mu v=v^a\nabla_a(\ln \mu)$. In the
infinite dimensional case, this has to be properly reinterpreted
in terms of ``functional derivatives''. In our case, the
logarithmic derivative of the measure along the constant vector
field defined by $g$ yields a quantum momentum operator of the
form,
\begin{equation}
\hat{\pi}[g]\cdot \Psi[\varphi]=-i\int_{\Sigma}
\biggl(g  {{\delta}\over{\delta \varphi}}
-\varphi\,G\,g \biggr) \Psi[\varphi] \, , \label{momalg}
\end{equation}
This completes the kinematical quantization from the algebraic
point of view, for a general Gaussian measure $\mu_{\rm G}$
\cite{draft}. Note that `correction term' in (\ref{momalg}) is the
most natural choice (from the infinite possibilities that the
algebraic approach allows) for the momentum operator.

{\it GNS Construction.} Let us now briefly review the results of
\cite{ccq}. The Fock representation of a scalar field on an
arbitrary space-time is based on the construction of a one
particle Hilbert space $\H_{\rm 1-p}$, constructed by defining a
Hermitian inner product on the phase space $\G$. In turn, this
inner product can be constructed using a complex structure $J$ on
$\G$, together with the naturally defined Symplectic structure on
$\G$: $\mu_{\G}(\cdot,\cdot)=-\Omega(\cdot,J\cdot)$ \footnote{Note
that we are using $\mu$ for two different constructs: $\mu_{\G}$
represents  an inner product on phase space, and $\mu_{G}$
represents a Gaussian measure on the quantum configuration
space.}. When the inner product $\mu_{\G}$ satisfies certain
conditions, then the resulting Hilbert space can be used to
construct the Fock space \cite{wald}. In the canonical picture,
the complex structure has the form $-J_{\Gamma}(\varphi , \pi)=(A
\varphi + B \pi ,C\pi +D\varphi)$, where $A,B,C,D$ are suitably
defined operators on the space of initial conditions, satisfying
conditions between them \cite{aa:am}.
 The inner product on $\G$ is then,
$$ \mu_{\Gamma}((\varphi_{1} , \pi_{1}),(\varphi_{2}, \pi_{2}))=
\int_{\Sigma}\d^3x \,(\pi_{1} B \pi_{2} + \pi_{1} A \varphi_{2} -
\varphi_{1} D \varphi_{2}-\varphi_{1} C \pi_{2} ).$$

The idea of the GNS construction is to start with the Weyl algebra
generated by the elements of the form
$\hat{W}[\lambda]=\exp(i\hat{F}[\lambda])$ and define an {\it
algebraic state} $\omega$ that maps the Weyl algebra to the
complex number.  The GNS theorem assures us that there is a unique
(up to unitary equivalence) representation of the CCR on a Hilbert
Space, such that the action  of the state $\omega$ on the
generators of the Weyl generators can be understood as the vacuum
expectation values of the Weyl operators.

The key step in the construction of the Schr\"odinger-Fock
representation, is to use the algebraic state that depends on the
phase space inner product $\mu_{\G}$. The algebraic state on an
arbitrary curved spacetime then takes the form \cite{ccq}
  \beq
   \omega_{\rm
fock}(\hat{W}(\lambda))=e^{-\frac{1}{4}\mu_{\G}(\lambda,\lambda)}
\, . \label{magia} \eeq

Then, by imposing this condition on the basic generators
$\hat{W}(\lambda)$ (corresponding to the exponentiated versions of
the configuration $\varphi[f]$ observables), one finds that the
quantum measure is of the form,
 \beq
\d\mu=e^{-\int_{\Sigma}\varphi B^{-1}\varphi}\;{\cal D}\varphi \,
. \label{medida}
 \eeq
Note that the measure knows only about one of the operators
defining $J_{\G}$, namely $B$. One expects, however that the full
quantum theory knows about at least another of the operators in
$J$ (two of which are independent). This expectation is realized
in the representation of the momentum operator.

The strategy is to assume a general form for the momentum operator
(\ref{momop}); one then applies the condition (\ref{magia}) to the
momentum operator $\hat{\pi}[g]$ and makes use of the
Baker-Campbell-Hausdorff relation. Then, it is straightforward to
show that the momentum operator takes the form,
 \beq
\hat{\tilde{\pi}}[g]\cdot \Psi[\varphi]=-i\int_{\Sigma} \biggl(g
{{\delta}\over{\delta \varphi}} -\varphi(B^{-1}+iB^{-1}A)g \biggr)
\Psi[\varphi] \, , \label{momccq}
\eeq
 where $B^{-1}$ and $B^{-1}A$ are also  Hermitian operators, and $B^{-1}$
is the operator corresponding to $G$ in (\ref{momalg}). Note that
the momentum operators $\hat{\tilde{\pi}}[g]$ and
$\hat{\tilde{\pi}}[h]$ commute for any choice of $g,h$ given that
$B^{-1}A$ is self adjoint. The most notorious fact about the
momentum operator thus found (\ref{momccq}) is that it has an {\it
extra term}, as compared to (\ref{momalg}).

A possible worry about this term is that the operator might not
satisfy the required reality conditions.  However, the extra term,
$\left[\int_{\Sigma} \varphi (B^{-1}A)\,g\right] \;
\Psi[\varphi]$, being a configuration operator smeared with a real
test function $B^{-1}A\,g$ is already a Hermitian operator. Thus,
the reality conditions are satisfied in both representations
(\ref{momalg}) and (\ref{momccq}). Note also that from the
geometric point of view, the representation found with the
algebraic perspective (\ref{momalg}) is the most natural one, when
one has a cotangent bundle structure for the phase space and one
is working in the configuration representation (vertical
polarization). It is indeed intriguing that one has an alternative
and not so intuitive representation. What is then the relation
between them? In order to explore this question let us make in the
next section a brief detour into ordinary quantum mechanics where
this can be easily understood. We shall come back to this question
at the end of Sec.~\ref{sec:4}.


\section{Unitary Transformations}
\label{sec:4}


Let us take a mechanical system whose phase space $\G$ is given by
pairs $(q^i,p_j)$. The canonical Poisson brackets
$\{q^i,p_j\}=\delta^i_j$ induce, via the Dirac prescription, the
CCR $[\hat{q}^i,\hat{p}_j]=i\hbar\delta^i_j$. Now, let us suppose
we have two different representations of the CCR on the {\it same}
Hilbert space ${\cal H}=L^2(q,\d q)$, given by \beq
(\hat{q}^i\cdot\psi)(q):=q^i\,\psi(q)\quad ;\quad
(\hat{p}_j\cdot\psi)(q):=-i\hbar\frac{\partial}{\partial
q^j}\,\psi(q) \label{rep1} \eeq and \beq (\hat{\tilde
q}^i\cdot\psi)(q):=q^i\,\psi(q)\quad ;\quad (\hat{\tilde
p}_j\cdot\psi)(q):= \left(-i\hbar\frac{\partial}{\partial
q^j}+q^j\right)\, \psi(q)\, .\label{rep2} \eeq
The question we
want to ask is the following: Is there a way of relating both
representations via a {\it unitary transformation}? As we shall
see in the following, the answer is in the affirmative. Let us
recall that given a real function $T$ on the classical phase
space, the one parameter family of diffeomorphisms generated by
its Hamiltonian vector field $X^a_T$ induces a map $U(t)$ on the
algebra of functions on the phase space $f\rightarrow\,U(t)\cdot
f$ given by \cite{arlen,AA:TT}:
\begin{eqnarray}
U(t)\cdot f &=& \sum^\infty_{n=0}
\frac{{t}^n}{n!}\{f,T\}_n\nonumber\\
&:=& f+t\{f,T\}+\frac{t^2}{2!}\{\{f,T\},T\}+\cdots
\end{eqnarray}
One should note that for each value of $t$ the Poisson structure
of the phase space is preserved, inducing an automorphism on the
algebra of observables. The corresponding quantum operator
$\hat{U}$ should  then be an automorphism in the algebra of
quantum operators, and therefore, a unitary operator. It is clear
also that it should be of the form
$\hat{U}=\exp\left[\frac{i}{\hbar}\hat{T}\right]$, where $\hat{T}$
is Hermitian. Given a state $|\psi\rangle$, by means of the
operator it gets mapped to $\hat{U}|\psi\rangle$ and any
 observable $\hat{{\cal O}}$ will go to
$\hat{U}\cdot\hat{\cal O}\cdot\hat{U}^{-1}$. In the example
outlined above, it is clear that the generating function $T$ is
given by
\[
T=\frac{1}{2}\,\left(\sum_i q^iq^i\right)
\]
Therefore the automorphisms generated on the functions $F(q,p)=
\alpha_iq^i+\beta^jp_j$ are \beq F(q,p)\rightarrow
F(q,p)+\beta^jq_j \eeq thus, the elementary observables get mapped
as \beq \tilde{q}^i=q^i\quad ;\quad \tilde{p}_j=p_j+q_j\, . \eeq
The corresponding quantum unitary operator $\hat{U}$ is given by
\beq \hat{U}=\exp\left[\frac{i}{2\hbar}\sum_i
\hat{q}^i\hat{q}^i\right] \, ,\label{repqm2} \eeq
 and maps the corresponding operators as follows:
\beq \hat{\tilde{q}}^i=\hat{q}^i\quad ;\quad
\hat{\tilde{p}}_j=\hat{p}_j+\hat{q}_j \, .\label{repqm} \eeq This
completes out detour into quantum mechanics.

 Let us now return to
our field theory problem. We have two representations given by
(\ref{momalg}) and (\ref{momccq}) and note that they have
precisely the structure of (\ref{repqm2}) and (\ref{repqm}) ,
where the momentum operators differ by a pure configuration
observable. Therefore, the general theory of classical and quantum
canonical transformations as discussed above should also apply to
this system. It is straightforward to find the generating function
$T$ in the field theory case: \beq
T[(\varphi,\pi)]=\frac{1}{2}\int_\Sigma\d^3\!x\,\varphi\,K\,\varphi
\eeq where $K=B^{-1}A$. Therefore it is immediate to write down
the corresponding would-be generator of unitary transformations,
\beq
 \hat{U}=\exp\left[\frac{i}{2\hbar}\int_\Sigma
\d^3\!x\widehat{(\varphi K\varphi)} \right]\, .\label{unitary}
 \eeq
  Then, one might
be lead to conclude that the representations (\ref{momalg}) and
(\ref{momccq}) are unitarily equivalent, and therefore the
algebraic procedure provides us with a correct and equivalent
quantum theory to the representation found in \cite{ccq}.

Interestingly, this is indeed {\it not} the case. That is, the two
representations are, in general, unitary inequivalent. Let us now
try to see why this is the case. As we have briefly explained, the
quantization of the scalar field used in Ref.~\cite{ccq} that
yields the representation (\ref{momccq}) is based on the GNS
construction of a representation of the canonical Weyl algebra,
equivalent to a given Fock representation. The representation
(\ref{momccq}) was obtained asking that the {\it Weyl} operators
$\hat{W}[\lambda]=\exp(i\hat{F}[\lambda])$ be unitary and whose
vacuum expectation values coincide with those of the Fock
representation (that knows about both operators $G$ and $K$).
Then, the general representation should contain both terms in
(\ref{momccq}). Let us now consider the vacuum expectation values
of the resulting quantum theories. For the GNS construction, the
formula (\ref{magia}) yields,
 \beq
\langle\hat{W}(\lambda)\rangle_{\rm
gns}=\exp\left[-\frac{1}{4}\int_\Sigma\d^3\!x(fBf+fAg-gDg-gCf)\right]
\label{vac1} \eeq
 for a ``label vector" $\lambda^a=(g,f)^a$. On the other hand,
the representation (\ref{momalg}) will have, as vacuum expectation
values,
 \beq \langle\hat{W}(\lambda)\rangle_{\rm
alg}=\exp\left[-\frac{1}{4}\int_\Sigma\d^3\!x(fBf-gB^{-1}g)\right]
\label{vac2}
 \eeq
 From the previous expressions, it is clear that the absence
  of the extra term in the
representation (\ref{momalg}) implies that, in general, the vacuum
expectation values (\ref{vac1}) and (\ref{vac2}) differ. Thus, the
theories are unitarily inequivalent. This is the main observation
of this letter.

It is important to stress that the main point that we want to
emphasize here is the fact that theories that heuristically should
be equivalent, turn out not to be. That is, one might think that
under appropriate regularization procedures (given that one has
product of operators), one might be able to find a unitary
operator for $\hat{U}$ in (\ref{unitary}) on the Hilbert space of
the theory. The general results of \cite{ccq} tell us, however,
that for certain operators $K$, no such regularization exists. The
precise conditions under which both representations are unitary
equivalent will be reported elsewhere.



\section{Discussion and Outlook}
\label{sec:5}


In this letter we have analyzed the canonical quantization of a
real scalar field. We have shown that the quantization found by a
direct application of the algebraic approach \cite{AT}, using the
most simple and natural choice for the representation of the CCR
is incomplete. That is, a crucial term in the representation,
which is completely un-natural from the geometric-algebraic
viewpoint, is missing. Furthermore, this term seems to be, at
least at a heuristic level, unitarily implementable. However, more
rigorous results coming from GNS tell us that such a unitary
transformation does not exist in general. In a sense, these
results illustrate a class of ambiguities in the quantization of
field theories that were not explicitly considered before.

The algebraic approach to quantization is general enough that it
allows for `any'  representation of the CCR in the Schr\"odinger
picture. In this regard, the Fock-Schr\"odinger representation of
\cite{ccq} lies within this general scheme. However, what we have
shown is that there are, in fact, an infinite number of
inequivalent representations, labelled by $K$, that share the same
Hilbert space structure (i.e. measure $\mu_{\rm G}$). We have also
shown that the often used strategy of fixing the representation by
means of the measure misses this infinite freedom. Thus, one needs
some other principle to arrive at the correct representation for
the system under consideration.

Let us end with three remarks:

\begin{enumerate}

\item This `negative' result, together with recent results
concerning the unitary implementability of (arbitrary) time
evolution \cite{mad,ccq2} point in the direction that finite
`exponentiated' versions of (possibly) well defined Hermitian
operators might not in general exist as  operators in a rigorously
defined quantum field theory. Needless to say, a deeper
understanding of these features is needed.

\item In the literature it is sometimes assumed that the correct
representation for the momentum operator comes from
Eq.~(\ref{momalg}). This assumption is not completely unjustified,
since for the cases that are normally considered in the literature
--such as Minkowski and static space-times, for instance-- the
operator $K$ vanishes and therefore, both representations
coincide. It is only when one considers the most general case on
curved spacetimes that such a term manifests itself.

\item In the non-perturbative quantization of general relativity
\cite{loop}, where one needs to define a background free quantum
field theory, it is very important to have full control over the
whole realm of possible ambiguities in the quantization process.
The ambiguity pointed out in this note might be of some relevance
for the choice of right representation of the `electric field'
momenta operators in the quantum geometry formalism \cite{hanno},
and for analyzing issues related to the semi-classical, low energy
limit of the theory.

\end{enumerate}

\begin{acknowledgments}

We would like to thank A. Ashtekar for a careful reading of the
manuscript. This work was in part supported by DGAPA-UNAM grant.
No. IN112401, by CONACyT grants No. J32754-E and 36581-E and by
the UNAM-CONACyT Graduate program (J.C.)

\end{acknowledgments}

\end{document}